\documentclass{aa}
\usepackage{graphicx,epsfig,fancyhdr,psfig,rotating,amsmath,longtable}
\usepackage{natbib}

\bibpunct{(}{)}{;}{a}{}{,} 
  \newcommand{\Msolar}{\mbox{\,$\rm M_{\odot}$}}        
  \newcommand{\Rsolar}{\mbox{\,$\rm R_{\odot}$}}        
  \newcommand{\Teff}{\mbox{\,\em T$_{\rm eff}$}}        
  \newcommand{\logg}{\mbox{\,log $g$}}                  

  \newcommand{\eg}{\mbox{e.g.}}                          
  \newcommand{\etal}{\mbox{et~al.}}                      
  \newcommand{\ang}{\,\mbox{\AA}}                        
  \def\simge{\mathrel{\raise1.16pt\hbox{$>$}\kern-7.0pt
    \lower3.06pt\hbox{{$\scriptstyle \sim$}}}}           
  \def\simle{\mathrel{\raise1.16pt\hbox{$<$}\kern-7.0pt
    \lower3.06pt\hbox{{$\scriptstyle \sim$}}}}           
\def\fm{\hbox{$.\!\!^{\rm m}$}}

\begin{document}

\title{Discovery of magnetic fields in central stars of planetary nebulae\thanks{Based
on observations collected at the European Southern Observatory,
Paranal, Chile, under programme ID 072.D-0089}}
\author{Stefan Jordan\inst{1}
        \and Klaus Werner\inst{2}  \and Simon J. O'Toole\inst{3}
       }
\offprints{S. Jordan, \email{jordan@ari.uni-heidelberg.de}}
\institute{
Astronomisches Rechen-Institut, M\"{o}nchhofstr. 12-14, 69120 Heidelberg,
Germany
 \and
Institut f\"ur Astronomie und Astrophysik, Universit\"at T\"ubingen, Sand 1, 72076 T\"ubingen, Germany
 \and
Dr.-Remeis-Sternwarte Bamberg, Sternwartstr. 7, 96049 Bamberg, Germany
}

\date{Received 10 September 2004/ Accepted 1 November 2004}

\abstract{For the first time we have directly detected  magnetic fields in  central stars of
planetary nebulae by means of  spectro-polarimetry with FORS1 at the VLT. In all four objects of our sample
we found kilogauss magnetic fields, in  NGC\,1360 and LSS\,1362  with very high significance, while
in  EGB\,5 and Abell\,36 the existence of a  magnetic field is probable but with less certainty.
This discovery supports the hypothesis that the non-spherical symmetry of most
planetary nebulae  is caused by  magnetic fields in AGB stars. Our high discovery rate demands mechanisms to prevent full conservation of magnetic flux
during the transition to white dwarfs.
}

\authorrunning{Stefan Jordan \etal}
\titlerunning{Discovery of  magnetic fields in central stars of planetary nebulae
}

\maketitle

\section{Introduction}
The reason why more than 80\%\ of the known planetary nebulae (PNe) are mostly
bipolar and not spherically symmetric
\citep{Zuckerman-Aller:86,Stanghellini-etal:93}
is barely understood.
A popular explanation is the interacting stellar winds model
\citep{Kwok-etal:78}, where the fast
($v\approx  1000$\,km/sec, mass loss $\approx  10^{-7} \Msolar/$yr) wind from the central star of a
PN encounters an older slow ($v \approx  10$\,km/sec) wind from earlier phases with heavy mass loss
($\approx 10^{-5} \Msolar$/yr). The visible PN is formed in the shock region between both winds;
if the  slow wind was not spherical, but densest in the equatorial plane, the nebula is bipolar.
However, neither is this model indisputable, nor is the physical mechanism for the asymmetry of the
slow wind clear. One possibility is the presence of a low-mass companion star which could exert a
gravitational pull on the circumstellar envelope. Rapid rotation and
binarity \citep[e.g.\ ][]{DeMarco-etal:04} may also cause asymmetries, but
the most promising explanations involve magnetic fields.
There is, however, no agreement about the detailed mechanism.
A review on observational and theoretical studies of the shaping of
planetary nebulae is given by \cite{Balick-Frank:02}.

It is possible that magnetic fields from the stellar surface are wrapped up by differential rotation so that
the later post-AGB wind will be collimated into two lobes
\citep{Garcia-etal:99}. Another scenario
says that  magnetic pressure at the stellar surface plays an important role driving the stellar
wind on the AGB \citep{Pascoli:97}.

The idea that  magnetic fields are important has been supported by the detection of polarization in
radio data of circumstellar envelopes of AGB stars: SiO (at a distance of   5-10 AU from the star),
H$_2$O ($\approx  100$ AU), and OH (100-1000 AU)  masers
\citep{Kemball-Diamond:97,Szymczak-Cohen:97,Vlemmings-etal:02}.

For H$_2$O masers \cite{Vlemmings-etal:02} are convinced that the  Zeeman interpretation is correct
and that the magnetic field strength at the H$_2$O maser of the Mira variable U Her is about 1.5G.
Depending on the topology of the magnetic field, the corresponding surface magnetic field is of the
order of 100-1000 G.

The magnetic field may be either a fossil remnant from the progenitor on the
main sequence (e.g. Ap stars), or can be generated by a dynamo at the
interface between a rapidly rotating stellar core and  a more slowly rotating
envelope.  \cite{Blackman-etal:01}  argue that some remnant field anchored in
the core will survive even without a convection zone, although the convective
envelope may not be removed completely. \cite{Thomas-etal:95} have shown that
white dwarfs which do have thin surface convection zones can support a
near-surface dynamo. Since the field strength in
their model is higher at higher luminosities this would particularly be true
for central stars of PNe.

That  some central stars must contain significant magnetic fields is also obvious from the fact
that at least  10-30\%\ of all white dwarfs have magnetic  fields  between  $10^3$ and $10^9$ Gauss.
Until now no magnetic fields have ever directly been detected in central stars of PNe.

We have observed a sample of four central stars of planetary nebulae with high signal-to-noise (circular)
spectropolarimetry between 3500 and 5900\,\AA\  with the FORS1 spectrograph of the VLT telescope.
As was already demonstrated for bright white dwarfs \citep{Aznar-etal:04},
the unprecedented light collecting power
of the VLT offers  the possibility to investigate
the presence of magnetic fields on the kG level.

\section{Observations and data reduction}
The observations  were obtained
in service mode between
November 2, 2003, and January 27, 2004, with the FORS1 spectrograph
of the UT1 (``Antu'') telescope of the  VLT, which is able to measure
circular polarization with the help of a
 Wollaston prism and rotatable retarder
 plate mosaics in the parallel beam allowing linear and circular polarimetry
 and spectropolarimetry \citep{Appenzeller-etal:98}.
We used grism G600B, covering the spectral range 3400--5900 \ang,
and a 0.8$\arcsec$ wide slit, leading to a spectral resolution
of 4.5\ang.
The details of our observations  are listed in
Table\,\ref{t:obs}.
The four selected objects  are bright ($V\le 12\fm 5$) and their nebulae
show clear indications for non-spherical symmetry.
So that we obtained  at least one good result for a star, we decided
to spend four times as much observing time on the CPN of NGC\,1360
then on the other three objects, for which only one observing block
was performed.
In order to reduce errors from changes in the sky transparency,
atmospheric scintillation, and various instrumental effects
the  $\lambda/4$-retarder plate was rotated by  $90^\circ$
after $n$ exposures (where $n$ is given in the last column of Table
\ref{t:obs}). The same number of exposures were then taken in this
configuration.

\subsection{Data reduction}
Calibration frames (bias, flat-field and He+HgCd arc spectra) were taken
during the day, following each nights observations.
The data were reduced in the \textsc{iraf} environment using the
following procedure. The bias level was subtracted from all frames and cosmic
rays were removed. A nightly master flat field was then constructed
from each night's individual flat fields. After flat-field correction, the
stellar spectra were extracted from each frame by summing up all CCD rows for
the ordinary and extraordinary ($e$ and $o$) beams. Background sky light was averaged over 10 rows (giving
a total of 20 rows) on either side of the object spectrum and subtracted. It
is important to note that the automatic aperture and sky selection routine in
\textsc{iraf} does \emph{not} always use the user-defined values, so each
spectrum was checked manually.

Wavelength calibration is particularly important for this kind of
spectropolarimetric study, and special care was taken to ensure its
accuracy. Failure to do so would lead to spurious polarization signals in
every line. Calibration was done independently for the spectra of each beam
and each position of the retarder plate (i.e.\ the $e$ and $o$ beams at $\pm
45^\circ$). 

 The referee suggested that spurious signals may be caused by using arc spectra
taken at different waveplate angles, probably because the spectra are
rebinned differently. To test this we examined two cases: when the
dispersion correction was applied, all spectra were forced to exactly the same
scale or they were simply corrected according to the dispersion function
only. In these two cases the spectra were rebinned differently. When we
determine the magnetic field strength, however, the results are the same
within errors. This indicates that rebinning is not affecting our results.
We have also examined the sky spectral lines at the edge of our
spectra; these lines show no detectable polarisation, suggesting that any
polarisation we measure is intrinsic to the star and not due to poor
wavelength calibration. Finally we note that while instrumental polarisation
dominates the Stokes V/I spectrum when considering only one waveplate angle,
we are encouraged to see the polarisation profiles at the positions of the
Balmer and He\,\textsc{II} lines.
The wavelengths are accurate to typically $\sim$3\,km\,s$^{-1}$ or
$\sim$0.05\,\AA\ at H$\beta$. This is much lower than the spectral resolution.

\begin{figure*}[tbhp]
\epsfxsize=16.cm \epsfbox{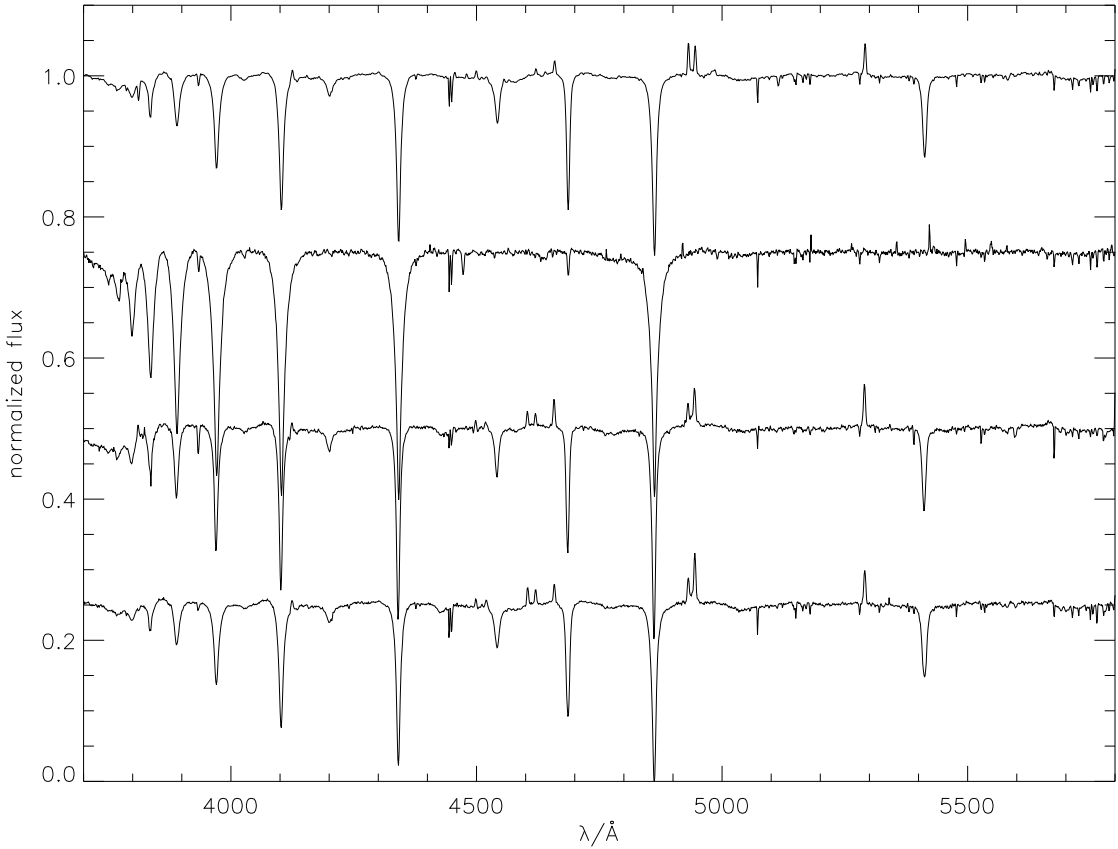}
\caption{Normalized spectra of our sample of central stars of planetary
nebulae (from above displaced vertically: NGC\,1360, EGB\,5, LSS\,1362, Abell\,36)
}
\label{f:ni}
\end{figure*}

\begin{figure*}[tbp]
\centerline{\psfig{figure=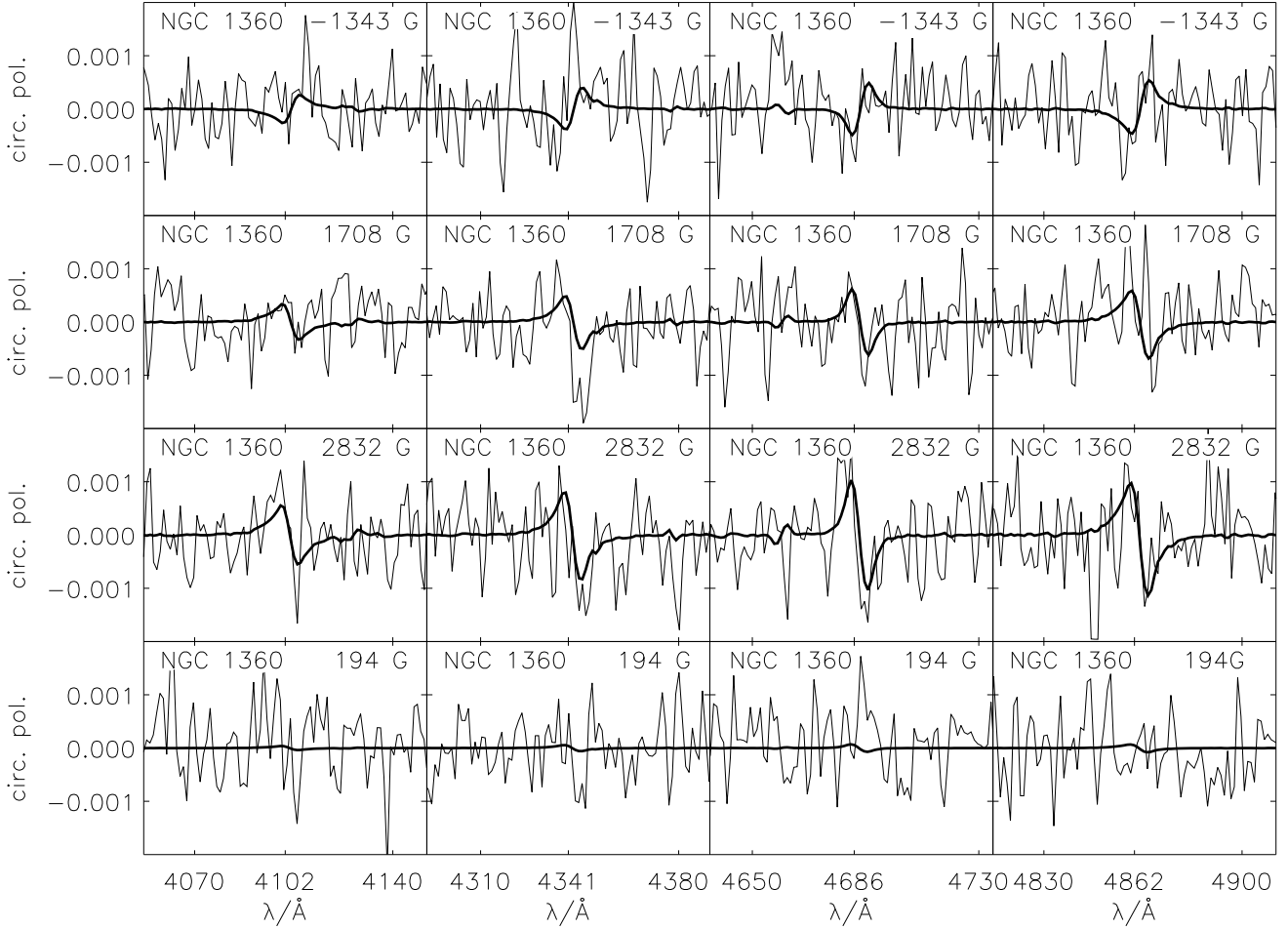,width=\textwidth}}
\caption{Circular polarization  ($V/I$) in the four  observation
blocks of 
of the central star of NGC\,1360 in the vicinity of the strong
spectral lines H$\delta+$He\,\textsc{II}, H$\gamma+$He\,\textsc{II}, He\,\textsc{II} 4686, H$\beta+$He\,\textsc{II}
compared to the prediction by the low-field
approximation (Eq.\,\ref{e:lf}) using a longitudinal magnetic
field of $-1343$\,G, $1708$\,G, $2832$\,G, and $194$\,G, respectively}
\label{f:ngc}
\end{figure*}

\begin{figure*}[tbhp]
\centerline{\psfig{figure=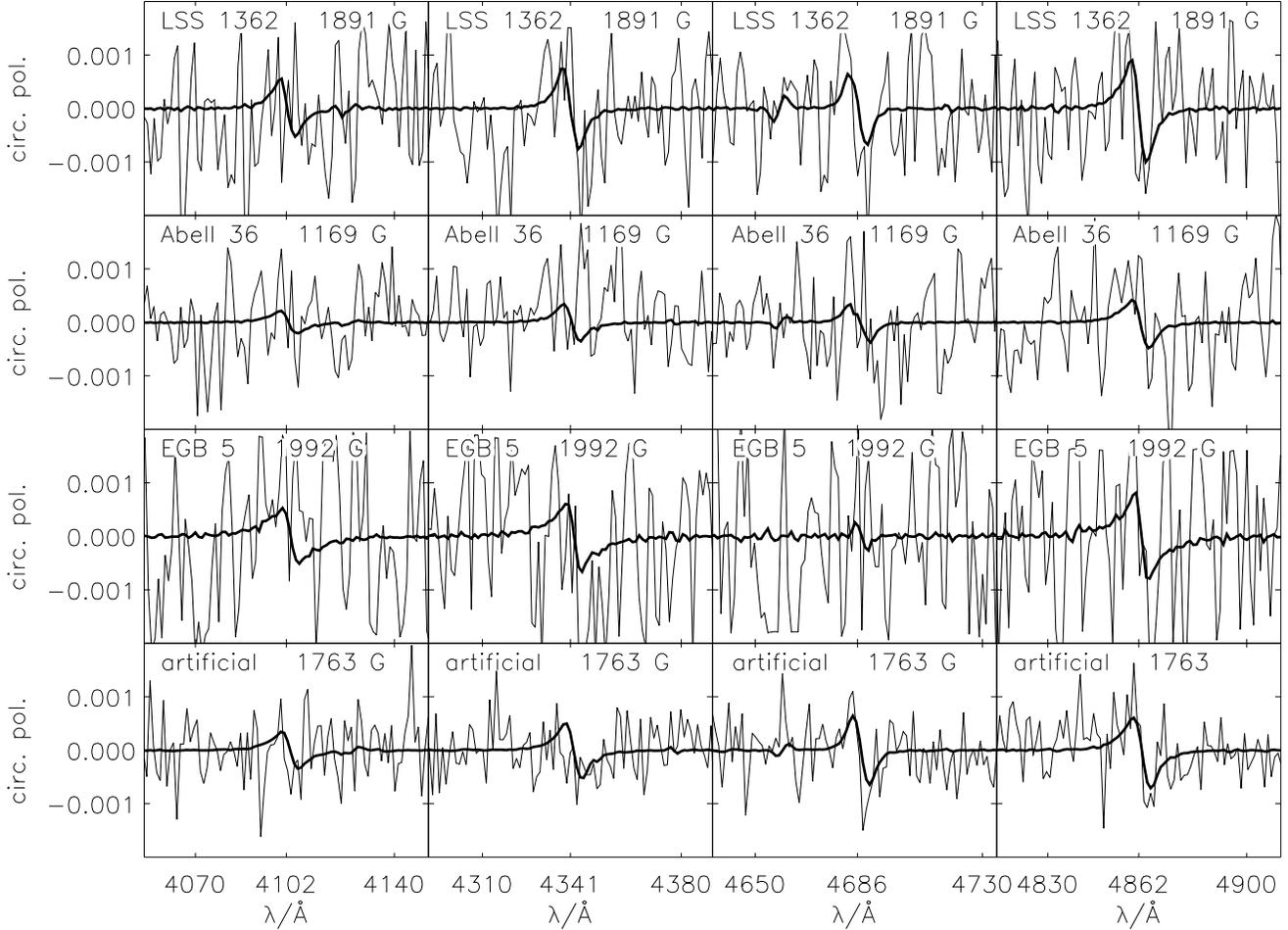,width=\textwidth}}
\caption{Circular polarization  ($V/I$) observed
in  the central stars  of LSS\,1362, Abell\,36, and EGB\,5
in the vicinity of the strong
spectral lines H$\delta+$He\,\textsc{II}, H$\gamma+$He\,\textsc{II}, He\,\textsc{II} 4686, H$\beta+$He\,\textsc{II}
compared to the prediction by the low-field
approximation (Eq.\,\ref{e:lf}) using a longitudinal magnetic
field of $1891$\,G, $1169$\,G, and $1992$\,G.
In the lower level we show an example for a fit to one of the
artificial spectra with a noise level of 0.0005 and an
assumed magnetic field of 1500\,G. The fit results in
$B=1763$\,G}
\label{f:other}
\end{figure*}

Stokes $I$, or unpolarized, spectra were obtained simply by summing all
spectra taken of an object in a single night. The Stokes $V/I$ spectra,
describing the net circular polarization, were created by summing the
exposures made at the same retarder plate position angle, and then applying
the following equation
\begin{equation}
\frac{V}{I}=\frac{R-1}{R+1},\; \mathrm{with}\: R^2=\left (\frac{f_o}{f_e}
\right )_{\alpha=+45} \times \left (\frac{f_e}{f_o} \right )_{\alpha=-45},
\end{equation}
which is equivalent to  formula 4.1 in
 the FORS 1+2 User Manual \citep{Szeifert-Boehnhardt:03}.
Here $\alpha$ indicates the nominal value of the position angle of the
retarder-wave plate, and $f_o$ and $f_e$ are the fluxes on the detector from
the $e$ and $o$ beams of the Wollaston prism, respectively.
The resulting high-quality spectra are shown in Fig.\,\ref{f:ni}, while
Figs.\,\ref{f:ngc}$-$\ref{f:other} show the circular polarization ($V/I$)
spectra.

\begin{table*}[tbp]
\caption{Details of VLT observations. The coordinates  $\alpha$ and $\delta$
refer to epoch 2000.
}
\begin{center}
\label{t:obs}
\begin{tabular}{cccccccc}
\hline
\hline
Target & Alias & $\alpha$ & $\delta$ & $V$ & HJD & $t_{\mathrm{exp}}$ & n \\
 & & & & (mag) & (+2\,452\,900) & (s) & \\
 \hline
 NGC\,1360 & CD--26 1340 & 03 33 14.7 & $-$25 52 18 & 11\fm 2 & 46.782 & 104 & 6 \\
  & & & & & 88.725 & 104 & 6 \\
   & & & & & 89.549 & 104 & 6 \\
    & & & & & 90.571 & 104 & 6 \\
    EGB\,5 & PN\,G211.9+22.6 & 08 11 12.8 & +10 57 19 & 12\fm 5 & 88.854 & 331 & 3 \\
    LSS\,1362 & PN\,G273.6+06.1 & 09 52 44.5 & $-$46 16 51 & 12\fm 5 & 89.796 & 331 & 3 \\
    Abell 36 &PN\,G318.4+41.1  & 13 40 41.4 & $-$19 52 55 &11\fm 5 & \llap{1}31.773 & 150 & 5 \\
    \hline
    \end{tabular}
    \end{center}
    \end{table*}

\section{Determination of the  magnetic field strenghs}
For weak magnetic fields (i.e. below 10\,kG) theoretical polarization
spectra ($V/I$) can be obtained by using the
weak-field approximation
\citep[\eg,][]{Angel-Landstreet:70,Landi-Landi:73}:
\begin{equation}
\frac{V}{I} = -g_{\rm eff} \ensuremath{C_z}\lambda^{2}\frac{1}{I}
\frac{\partial I}{\partial \lambda}
\ensuremath{\langle\large B_z\large\rangle}\;,
\label{e:lf}
\end{equation}
where $g_{\rm eff}$ is the effective Land\'{e} factor
\citep[which is unity for Balmer lines and for the
hydrogenic He\,\textsc{II} lines;][]{Casini-Landi:94},
$\lambda$ is the wavelength expressed in \ang, $\langle B_z\rangle$ is the
mean longitudinal component of the magnetic field  expressed in Gauss and the
constant
$C_z=e/(4\pi m_ec^2)$ $(\simeq 4.67 \times10^{-13}\,{\rm G}^{-1} \ang^{-1})$.
Since we do not have any information about the detailed field geometry
we can only measure the mean longitudinal field over the stellar
surface. The maximum
field strength can be larger than this value.

Since both the hydrogen lines and the He\,\textsc{II} lines have an
effective Land\'{e} factor of unity we do not expect blending to have a large influence. However, it is clear that the total effect of two
separate spectral lines on the polarization is not the same
as treating a blended line in the same way. With our
method it is not possible to disentangle both effects. Test
calculations using theoretical spectra for NGC\,1360 have
shown that the result when using blended Balmer and
He\,\textsc{II} lines instead of a sum of non-blended lines (by
switching hydrogen or helium, respectively, in the calculation
of the theoretical spectrum) differ by only about
200\,G.

\begin{figure}
\epsfxsize=9cm \epsfbox{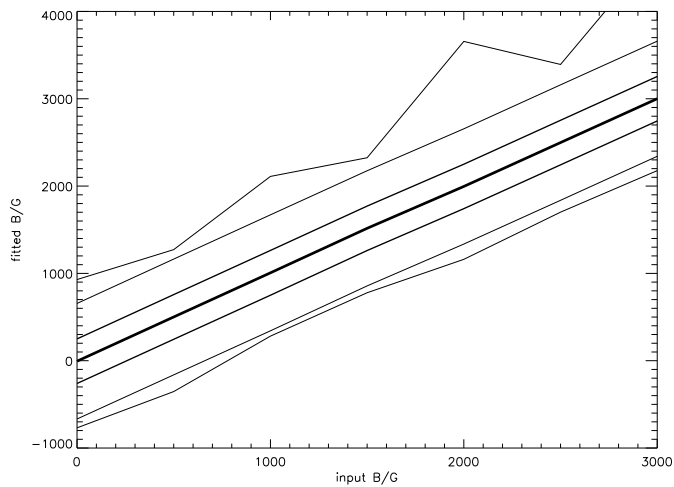}
\caption{Result of the fits with simulated data having the
same noise level as the observations of NGC\,1360 for
input magnetic fields between 0 and 3000\,G, in steps of
500\,G. From center line to
outside: mean fit result, 1$\sigma$ error range, 99\%\ confidence level,
and smallest and largest fit result during the 1000 simulations which
were performed for each predescribed magnetic field.
}
\label{f:sim_ngc_acc}
\end{figure}


The
longitudinal component of the magnetic field  for
each measurement was determined by comparing
the observed circular polarization for an
interval of $\pm 20$\,\ang\ around the four strongest absorption lines H$\beta+$He\,\textsc{II},
He\,\textsc{II} 4686, H$\gamma+$He\,\textsc{II}, H$\delta+$He\,\textsc{II}
with the prediction of Equation\,\ref{e:lf}. As in \cite{Aznar-etal:04}
we determined $\langle B_z\rangle$
by a $\chi^2$-minimization procedure.
Following \cite{Press:86} we determined the statistical error from the rms
deviation of the observed circular polarization from the best-fit model.
The $1\sigma$ (68.3\%) confidence range for a degree of freedom of 1 is the
interval of $B_z$ where the deviation from the minimum is $\Delta \chi^2=1$;
the 99\%\ confidence interval corresponds to
$\Delta \chi^2=6.63$.
This statistical error does not take into account any systematic
errors, particularly the blending of Balmer lines with
He\,\textsc{II} lines mentioned above. Only the He\,\textsc{II} 4686 line is
not effected by blending.
Although not blended, the weaker He\,\textsc{II} lines do not give any
significant information; they have large statistical errors and therefore a
very low weight.

For each of the observation blocks,
Table\,\ref{t:mf} summarizes our fit results for all four spectral
lines and the weighted means
$B_z=(\sum B_{z,i} w_i)/\sum w_i$ with $i$ corresponding to the
lines and $w_i=1/\sigma_i^2$.
The total probable error is given by
$\sigma=(\sum w_i)^{-1/2}$. We list both the total $\Delta \chi^2=1$
 and
$\Delta \chi^2=6.63$ error range.
\begin{table*}[tbp]
\begin{center}
\caption{Magnetic fields derived from the four strongest lines in our
sample of central stars of planetary nebulae. The error margins
correspond to a $1\sigma$ (68.3\%\ confidence) and
$6.6\sigma$ (99\%\ confidence) level.
}
\label{t:mf}
\begin{tabular}[c]{lcrrrrr}
\hline
\hline
\multicolumn{1}{c}{Target} &
\multicolumn{1}{c}{Date} &
\multicolumn{4}{c}{$B$/G} &
\multicolumn{1}{c}{$B$/G}
\\
\multicolumn{1}{c}{}  &
\multicolumn{1}{c}{}  &
\multicolumn{1}{c}{H$\delta+$He\,\textsc{II}} & \multicolumn{1}{c}{H$\gamma+$He\,\textsc{II}} &
\multicolumn{1}{c}{He\,\textsc{II} 4686} & \multicolumn{1}{c}{H$\beta+$He\,\textsc{II}} &
\multicolumn{1}{c}{total}  \\
\hline
NGC\,1360 &03/11/03& $-493\pm 835$ & $2483\pm688$
&   $-1114\pm427$ &    $-1355\pm413$ &  $-1343\pm259$  \\
&& $(\pm 2153)$ & $(\pm 1772)$
&   $(\pm 1101)$ &    $(\pm 1066)$ &  $(\pm 668)$  \\[2pt]
NGC\,1360&14/12/03& $342\pm 624$ & $3553\pm528$
&   $1726\pm426$ &    $768\pm543$ &  $1708\pm257$  \\
&& $(\pm 1607)$ & $(\pm 1361)$
&   $(\pm 1099)$ &    $(\pm 1399)$ &  $(\pm 664)$  \\[2pt]
NGC\,1360&15/12/03& $3146\pm 735$ & $2082\pm695$
&   $3714\pm389$ &    $1324\pm557$ &  $2832\pm269$  \\
&& $(\pm 1895)$ & $(\pm 1790)$
&   $(\pm 1002)$ &    $(\pm 1437)$ &  $(\pm 695)$  \\[2pt]
NGC\,1360&16/12/03& $2548\pm 1024$ & $1303\pm505$
&   $-1176\pm591$ &    $316\pm438$ &  $194\pm277$  \\
&& $(\pm 2638)$ & $(\pm 1302)$
&   $(\pm 1522)$ &    $(\pm 1128)$ &  $(\pm 2548)$  \\[2pt]
EGB\,5&14/12/03& $577\pm 1171$ & $2875\pm1022$
&   $484\pm4707$ &    $2171\pm837$ &  $1992\pm562$  \\
&& $(\pm 3016)$ & $(\pm 2633)$
&   $(\pm 12121)$ &    $(\pm 2156)$ &  $(\pm 1449)$  \\[2pt]
LSS\,1362&15/12/03& $-299\pm 1295$ & $2089\pm729$
&   $3608\pm894$ &    $1550\pm531$ &  $1891\pm371$  \\
&& $(\pm 3335)$ & $(\pm 1878)$
&   $(\pm 1848)$ &    $(\pm 1368)$ &  $(\pm 912)$  \\[2pt]
Abell\,36&26/01/04& $1863\pm 1144$ & $-448\pm949$
&   $1842\pm1134$ &    $1553\pm719$ &  $1169\pm466$  \\
&& $(\pm 2946)$ & $(\pm 2444)$
&   $(\pm 2920)$ &    $(\pm 1852)$ &  $(\pm 1202)$  \\[2pt]
\hline
\end{tabular}
\end{center}
\end{table*}
From our statistic a significant magnetic field was found in
three of the four NGC\,1360 observations and in
the (single) observations of EGB\,5, LSS\,1362, and
Abell\,36. However, in the latter case the value of the
best fit is just outside the 99\%\ confidence range.

NGC\,1360 clearly shows the effect of rotation between
the observations: $-1343, 1708, 2832,$ and $194$\,G.
The  difference in time between the three observations
was 42, 0.8, and 1.0 days. 
\cite{Werner-etal:03}
have derived an upper limit for the rotational velocity of
20\,km/sec from the with of iron lines, leading to  a period
larger than 0.75 days for a radius of 0.3\,\Rsolar, which
is compatible with our result.
The successors of CPNs (white dwarfs) are also rotating slowly
\citep{Koester-etal:98}, so that we do not expect any smearing out of the polarization signal
during the observing blocks.

\subsection{Statistical significance of our measurements}
Since the amplitude of a polarization signal for $B\approx 1$\,kG
is usually smaller
than the $1\sigma$ noise level of the observed polarization spectra,
doubts about the significance of our result clearly originate
when visually looking at the fitted polarization spectra.
For this reason we have started a simulation using synthetic
polarization spectra to which  Gaussian noise of the same level as in our
observation was added.

In the case of NGC\,1360 the noise in the single observed polarization
spectra has $\sigma_{\rm noise}=0.0005$.
For  given magnetic fields of $0,\dots, 3000$\,G in steps
of $500$\,G,
we calculated 1000 artificial polarization ``measurements'' and
treated them in the same way as our real observations.

Figure \ref{f:sim_ngc_acc} shows that the averaged weighted
mean for the four strong spectral lines is very close to the
given value of the magnetic field. It also shows that
for an assumed magnetic field of $B=0$\,G, only one result reaches
900\,G.
If we conservatively assume that the systematic error is 500\,G,
two of the four observing  blocks of NGC\,1360 have a much larger
measurement (1708 and 2822\,G), and one (-1342\,G) is only marginally
below this extremely pessimistic criterion.
Of all the simulations, 99\%\ with
an assumed $B=0$\,G have fitted field strengths below 660\,G.
On the other hand, if we assume a magnetic field of 1000\,G,
the fits to the artificial spectra result in values between
$280$\,G and 2110\,G,  with 99\%\ of them lying between
342 and 1670\,G.

The lower panels in Fig\,\ref{f:other} show an example of one of the 1000 artificial
spectra for an assumed magnetic field of 1500\,G, which is close
to the 1708\,G value for NGC\,1360 measured from the second observing block.
It makes it clear  that,
as already demonstrated by \cite{Aznar-etal:04}, 
visual inspection is misleading, since the eye does
not take into account an average small excess of right- and left-handed
polarization on different sides of the line core, respectively, which
contributes to our $\chi^2$ analysis.
The standard deviation of all 1000 fits is 254\,G, very close to our
formal  $1\sigma$ error for 1708\,G, which is 258\,G.
We therefore conclude
that in the case of NGC\,1360 the statistical errors from our $\chi^2$ analysis are
indeed realistic in order to judge how accurately the magnetic
field can be determined.

A somewhat different situation occurs in the case of LSS\,1362
($B_{\rm fit}=1891$\,G),
where the noise level of $\sigma_{\rm noise}=0.00085$ is larger.
If we assume that no magnetic field exists,
four of the 1000 simulations result in a  fitted magnetic field strength
exceeding  1891\,G. If, in order to account for a possible systematic error,
we set the limit at 1500\,G, 16 (1.6\%) of the simulations
provide a larger field strength.
The standard deviation for an assumed field strength of 2000\,G
is 646\,G, about 75\%\ larger than the formal $1\sigma$ error
from the $\chi^2$ analysis.
Therefore the probability that
LSS\,1362 has a magnetic field of more than 1000\,G is very
high.

In the case of Abell\,36, where we measured a magnetic field of
$1169\pm 466$\,G, the situation is more uncertain: for
$\sigma_{\rm noise}=0.00067$ we find that for an assumed
magnetic field of 0\,G 144 (14.4\%) of all artificial
polarization spectra mimic a magnetic field $>1169$\,G,
555 (55.5\%) a magnetic field larger than 669\,G (if we
again estimate the maximum systematic error to be 500\,G).
Therefore, we would not regard the derived magnetic
field as very significant.

Although we formally measured a magnetic field of
$1992\pm 562$ in EGB\,5, the case for a kilogauss magnetic field
is probable but not with the high certainty indicated by the error
range from the $\chi^2$ analysis. For 0\,G and $\sigma_{\rm noise}=0.0012$
we find that 64 (6.4\%) models exceeded 1992\,G, and
142 (14.2\%) the limit of 1492\,G, taking into account systematic
uncertainties. Due to the higher measured value, this is a clearer
case than that of Abell\,36.

Our simulations with artificial polarization spectra clearly show that much
more realistic error estimations can be obtained compared to the
formal errors from the $\chi^2$ analysis. They show, however, that
our determinations of magnetic fields are significant in the
case of NGC\,1360 and LSS\,1362 even though the maximum polarization
signal does not exceed the noise level.

\section{Parameters of the target stars and nebulae}
Atmospheric parameters for the central stars of
NGC\,1360, Abell\,36, and LSS\,1362
were derived by \cite{Traulsen-etal:05} from HST STIS and optical  spectra.
For EGB\,5 $T_{\rm eff}$ and $\log g$ were
determined by \cite{Lisker-etal:04}.
From these values, masses and radii for the central stars were
estimated, as well as the masses and radii on the main sequence
and of the white dwarf successors. For this purpose, mass-radius
relations by \cite{Wood:94} were used. In the case of
EGB\,5 no such values could be derived, since its central star (a hot
subdwarf) is a result of binary evolution \citep{Karl-etal:03}.
The value of these parameters together with a designation of the
planetary nebular morphology is listed in Table\,\ref{tab_targets}.

If we assume complete conservation of magnetic flux through
the stellar surface from the
main sequence to the white dwarf stage, we can estimate the magnetic
field strength of the precursors and successors. The magnetic field
strength measured from the third observing block in the central
star of NGC\,1360 of 2800\,G would translate into a field strength
of 50\,G on the main sequence while the field strength will be  enhanced
to 2\,MG is the star will reach the white dwarf stage. For Abell\,36
(1170\,G)
and LSS\,1362 (1900\,G) the  values are 9.3\,G, 0.35\,MG,
24\,G, and 0.43\,MG, respectively.

This is surprising, because magnetic fields of 0.35-2.0\,MG would
be detectable from Zeeman splitting in high-resolution and high-signal-to-noise
spectra, e.g. from the SPY survey \citep{Napi-etal:03} and
in the majority of the sample stars such high magnetic field strengths
can be excluded. Therefore, we have to assume that our assumption of
full conservation of magnetic flux is invalid. This might be a
hint that the magnetic field is not strongly  concentrated to the
degenerate stellar core, where the time scale for the decay should
be of the order of $10^{10}$ years \citep{Chanmugam-Gabriel:72,
Fontaine-etal:73}. It could instead be present in the
envelope, where it might be destroyed by convection or mass-loss.

\begin{table*}[tbp]
\caption{Characteristics of our program stars and their nebulae. The last column
gives the references for stellar parameters and PN shape. The main-sequence
(MS) mass is inferred from Weidemann's (2000)  initial-final mass relation. The MS
radius is estimated using \cite{Allen:76}. The future white dwarf (WD) radius is
estimated from the mass-radius relation of \cite{Wood:94}. EGB~5 is not on a post-AGB
evolutionary track as a result of close-binary evolution \citep{Karl-etal:03}.
\label{tab_targets}}
\begin{tabular}{l r r r r r r r r r l l}
\hline
\hline
\noalign{\smallskip}
Object  & \Teff & \logg & mass      & MS mass   & radius    & MS radius & WD radius &PN shape& ref.\\
&  [K]  & (cgs) &[M$_\odot$]&[M$_\odot$]&[R$_\odot$]&[R$_\odot$]&[R$_\odot$]& &        \\
\noalign{\smallskip}
\hline
\noalign{\smallskip}
 NGC\,1360 &  97\,000 & 5.3   & 0.65      & 2.7       & 0.30      & 2.3  &  0.011    & elliptical & A, B\\
 Abell 36  & 113\,000 & 5.6   & 0.60      & 2.7       & 0.21      & 2.3  &  0.012    & irregular & A, B\\
 LSS\,1362 & 114\,000 & 5.7   & 0.60      & 2.0       & 0.18      & 1.6  &  0.012    & ellip.ring & A, C\\
 EGB 5     &  34\,000 & 5.8\rlap{5}   &   0.48\rlap{$^*$}    &        &   0.14   &         &       & elliptical & D\\
 \noalign{\smallskip}
 \hline
 \end{tabular}
  \\References:
A: \cite{Traulsen-etal:05},
B: \cite{Phillips:03},
C: \cite{Heber-etal:88},
D: \cite{Lisker-etal:04}

$^*$Canonical mass of a hot subdwarf is assumed.
\end{table*}

\section{Discussion and conclusions}
We have detected magnetic fields in 50\%-100\%\ of our small survey
for magnetic fields in central stars of planetary nebulae, depending
on how conservatively  the criteria for statistical significance are
set. This provides very strong support for theories which explain the
non-spherical symmetry (bipolarity) of the majority of planetary
nebulae by magnetic fields. In this first survey we have not performed
a cross check with any spherically-symmetric nebulae, although this
is planned as a follow-up.

Although based on only four objects, our extremely high discovery
rate demands that magnetic flux must be lost during the transition
phase between central stars and white dwarfs: if the magnetic flux
was fully conserved, our four central stars will have fields
between 0.35 and 2\,MG
when they become white dwarfs. Although the number of white dwarfs with
magnetic fields is still a matter of debate, with a range between about 3
and 30\%, even the latter value, which includes
objects with kG field strengths \citep{Aznar-etal:04}, is far
off our high number.
\cite{Liebert-etal:03} quantified the incidence of magnetism at
the level of $\sim$ 2\,MG or greater to be of the order of
 $\sim$10\%.
 This argument would not change by much
if we consider that we have so far only looked at central
stars with non-spherical symmetric nebulae.
An almost 100\%\ probability of magnetic fields larger that
100\,kG can be excluded by the data from the SPY survey \citep{Napi-etal:03} as
well as the sample from \cite{Aznar-etal:04}. It is also worth mentioning that
our central stars have typical white dwarf masses (0.48-0.65\Msolar) and are
not particularly massive.  White dwarfs with MG fields tend to be
more massive than non-magnetic objects \citep{Liebert:88}.

If the magnetic field is located deep in the degenerate
core of the central star, it is very difficult to imagine a
mechanism to destroy the ordered magnetic fields. Therefore,
it would be more plausible to argue that the magnetic field
in the central stars is present mostly in the envelope where it
can be affected by convection and mass-loss. For central stars
hotter than 100\,000\,K we do, however, not expect convection;
only in the central star of EGB\,5 we cannot exclude such a mechanism.

If we assume that the magnetic fields are fossil and magnetic
flux was conserved until the central-star phase, we
estimate that the field strengths on the main sequence
were 9-50\,G, which are not directly detectable. Therefore, our measurement
may indirectly provide evidence for such low magnetic fields on the main
sequence.

Polarimetry with the VLT has led to discovery of magnetic
fields in a large number of objects in the final stage
of stellar evolution: white dwarfs \citep{Aznar-etal:04},
hot subdwarf stars \citep{OToole-etal:05}, and now in central stars
of planetary nebulae. Although we have now provided a
good basis for the theoretical explanation of the planetary nebula
morphology -- which can more quantitatively be correlated
with additional observations in the future --  new questions about the number
statistics of magnetic fields in the late stages of stellar evolution have
been raised.

\acknowledgements{
We thank the staff of the ESO VLT for carrying out the service
observations. Work on magnetic white dwarfs in T\"ubingen is supported by 
DLR grant 50 OR 0201, SJOT by 
50 OR 0202. We thank the referee and Regina Aznar Cuadrado 
for valuable comments.}

\bibliographystyle{aa}

\begin{thebibliography}{}
\bibitem[Angel \& Landstreet (1970)]{Angel-Landstreet:70}
         Angel, J.R.P., \&\ Landstreet, J. D. 1970, ApJ, 160, L147
\bibitem[{Appenzeller et al.\ (1998)}]{Appenzeller-etal:98}
         Appenzeller, I., Fricke,  K., F\"{u}rtig,  W., \etal\ 1998,
         ESO-Messenger, 94, 1
\bibitem[Allen (1976)]{Allen:76}
         Allen, C.W. 1976, Astrophysical Quantities, London: The Athlone Press
\bibitem[{Aznar Cuadrado et al.\  (2004)}]{Aznar-etal:04}
         Aznar Cuadrado, R.,  Jordan, S., Napiwotzki, R.,
         Schmid, H.M.,   Solanki, S.K., \&\
         Mathys, G. 2004, A\&A,  423, 1081
\bibitem[{Balick \&\ Frank (2002)}]{Balick-Frank:02}
       Balick, B., \&\ Frank, A. 2002, Annu.Rev.Astron.Astrophys., 40, 439
\bibitem[{Blackman et al.\ (2001)}]{Blackman-etal:01}
         Blackman, E.G., Frank, A., Markiel, J.A., Thomas, J.H.,  \&\ Van Horn, H.M. 2001,
         Nature, 409, 485
\bibitem[{Casini \& Landi degl'Innocenti (1994)}]{Casini-Landi:94}
         Casini, R.,  \&\ Landi degl'Innocenti, E. 1994, A\&A, 291, 668
\bibitem[{Chanmugam \&\ Gabriel  (1972)}]{Chanmugam-Gabriel:72}
Chanmugam, G.,  \&\ Gabriel, M. 1972, A\&A, 16, 149
\bibitem[{De Marco  et al.   (2004)}]{DeMarco-etal:04}
De Marco, O., Bond, H.E., Harmer, D.,  \&\ Fleming, A.J. 2004, ApJ, 602, 93
\bibitem[{Ellis et al.\ (1984)}]{Ellis-etal:84}
 Ellis, G.L., Grayson, E.T., \& Bond, H.E. 1984, PASP, 96, 283
\bibitem[{Fontaine et al.\ (1973)}]{Fontaine-etal:73}Fontaine G., Thomas, J.H.,  \&\ Van Horn, H.M. 1973, ApJ, 184, 911
\bibitem[{Garc{\'i}a-Segura et al.\ (1999)}]{Garcia-etal:99}
         Garc{\'i}a-Segura, G., Langer, N., Rocyczca, M.F.,  \&\ Franco, J. 1999, ApJ, 517, 767
\bibitem[{Heber et al.\  (1988)}]{Heber-etal:88}
  Heber, U., Werner, K., \& Drilling, J.S. 1988, A\&A, 194, 223
\bibitem[{Karl et al.\   (2003)}]{Karl-etal:03}
   Karl, C., Napiwotzki, R., Heber, U., \etal\ 2003, in White
   Dwarfs, eds.\ D.\,de Martino, R.\,Silvotti, J.-E.\,Solheim, R.\,Kalytis,
   NATO Science Series II, Kluwer, Vol.\ 105, 43
\bibitem[{Kemball \& Diamond (1997)}]{Kemball-Diamond:97}
Kemball, A.J.,  \&\ Diamond, P.J. 1997, ApJ, 481, L111
\bibitem[{Koester et al.\ (1998)}]{Koester-etal:98}
         Koester, D., Dreizler, S., Weidemann, V.,  \&\ Allard, N. F. 1998,
         A\&A, 338, 612
\bibitem[{Kwok et al.\ (1978)}]{Kwok-etal:78}
Kwok, S., Purton, C.R.,  \&\ Fitzgerald, P.M. 1978, ApJ, 219, L125
\bibitem[{Landi degl'Innocenti \& Landi degl'Innocenti (1973)}]{Landi-Landi:73}
         Landi degl'Innocenti, E.,  \&\ Landi degl'Innocenti, M. 1973,
         Solar Phys., 29, 287
\bibitem[{Landstreet (1982)}]{Landstreet:82}
         Landstreet, J. D. 1982, ApJ, 258, 639
\bibitem[{Liebert (1988)}]{Liebert:88}
         Liebert, J. 1988, PASP, 100, 1302
\bibitem[{Liebert et al.\ (2003)}]{Liebert-etal:03}
         Liebert, J., Bergeron, P.,  \&\ Holberg, J.B. 2003, AJ, 125, 348
\bibitem[{Lisker et al.\   (2004)}]{Lisker-etal:04}
Lisker, T., Heber, U.,  \&\ Napiwotzki, R., \etal\ 2004, A\&A, in press
\bibitem[{Napiwotzki et al.\ (2003)}]{Napi-etal:03}
         Napiwotzki, R., Christlieb, N., Drechsel, H., et al. 2003,
         ESO-Messenger, 112, 25
\bibitem[{O'Toole et al.\  (2005)}]{OToole-etal:05}
O'Toole, S., Jordan, S., Friedrich, S.,  \&\  Heber, U. 2005, in White Dwarfs, eds.\
D.\,Koester, S.\,Moehler, ASP Conf.\ Series, in press
\bibitem[{Pascoli  (1997)}]{Pascoli:97}
Pascoli, G. 1997, ApJ, 489, 94
\bibitem[{Phillips (2003)}]{Phillips:03}
Phillips, J.P. 2003, MNRAS, 344, 501
\bibitem[{Press et al.\ (1986)}]{Press:86}
         Press, W.H., Flannery, B.P.,  \&\ Teukolsky, S.A. 1986,
         Numerical Recipes, Cambridge, Univ. Press.
\bibitem[{Stanghellini et al.\  (1993)}]{Stanghellini-etal:93}
Stanghellini, L., Corradi, R.L.M.,  \&\ Schwarz, H.E. 1993, A\&A, 279, 521
\bibitem[{Szeifert \&\ B\"ohnhardt (2003)}]{Szeifert-Boehnhardt:03}
          Szeifert, T., \&\  B\"ohnhardt, J.D. 2003, FORS1+2 User Manual 2.6;
          ESO document VLT-MAN-ESO-13100-1543
\bibitem[{Szymczak \& Cohen (1997)}]{Szymczak-Cohen:97}
Szymczak, M.,  \&\ Cohen, R.J. 1997, MNRAS, 288, 945
\bibitem[{Thomas et al. (1995)}]{Thomas-etal:95}
         Thomas, J.H., Markiel, A.,  \&\ Van Horn, H.M. 1995, ApJ, 453, 403
\bibitem[{Traulsen et al.\  (2005)}]{Traulsen-etal:05}
Traulsen, I., Hoffmann, A.I.D., Dreizler, S., Rauch, T., Werner,
K., \& Kruk, J.W. 2005, in White Dwarfs, eds.\
D.\,Koester,  \&\ S.\,Moehler, ASP Conf.\ Series, in press
\bibitem[{Vlemmings et al.\ (2002)}]{Vlemmings-etal:02}
Vlemmings, W.H.T., Diamond, P.J.,  \&\ van Langevelde, H.J. 2002, A\&A, 394, 589
\bibitem[{Weidemann (2000)}]{Weidemann:2000}
         Weidemann, V. 2000, A\&A 363, 647
\bibitem[{Werner et al.\ (2003)}]{Werner-etal:03}
         Werner, K., Deetjen, J.L., Dreizler, S., Rauch, T., \&\  Kruk, J.W. 2003,
	 in Planetary Nebulae: Their Evolution and Role in the Universe,
	 Proceedings of the 209th Symposium of the IAU,  Eds.  S. Kwok, M. Dopita, and R. Sutherland, p.169
\bibitem[{Wood  (1994)}]{Wood:94}
Wood, M.A. 1994, in The Equation of State in Astrophysics, IAU
  Coll.\ 147, eds.\ G.\,Chabrier, E.\, Schatzman, Cambridge University Press,
  p.\,612
\bibitem[{Zuckerman \& Aller (1986)}]{Zuckerman-Aller:86}
Zuckerman, B.,  \&\ Aller, L.H. 1986, ApJ, 301, 772
\end{thebibliography}

\end{document}